\documentstyle[12pt]{article}
\newlength{\abstractwidth}
\newcommand{\Tr}{{\rm Tr}\ }

\newcommand{\p}{\partial}

\newcommand{\zb}{\mathbb{Z}}

\def\bea{\begin{eqnarray}}
\def\eea{\end{eqnarray}}

\def\p{\partial}

\def\zb{\overline{z}}

\def\Cb{\overline{C}}

\def\Eb{\overline{E}}
\def\Nb{\overline{N}}
\def\Tb{\overline{T}}

\def\Ab{\overline{A}}
\def\Jb{\overline{J}}
\def\Qb{\overline{Q}}
\def\Rb{\overline{R}}
\def\12{\frac{1}{2}}
\def\t{\theta}
\def\bea{\begin{eqnarray}}
\def\eea{\end{eqnarray}}
\def\ba{\begin{array}}
\def\ea{\end{array}}
\def\be{\begin{equation}}
\def\ee{\end{equation}}

\begin{document}
\begin{titlepage}
\bigskip
\bigskip\bigskip\bigskip\bigskip
\bigskip \bigskip
\centerline{\Large \bf Tachyon Condensation of D2/D4-Brane System}
\bigskip \centerline{\Large \bf in Noncommutative Gauge Theory}

\bigskip\bigskip
\bigskip\bigskip
\centerline{Bin Chen$^{*}$\footnote{chenb@ictp.trieste.it}  and
Feng-Li Lin $^{\dagger}$\footnote{linfl@phya.snu.ac.kr}}
\bigskip
\centerline{\it ${}^*$High Energy Section} \centerline{\it the
Abdus Salam ICTP} \centerline{\it Strada Costiera, 11}
\centerline{\it 34014 Trieste, Italy}
\bigskip
\centerline{\it ${}^\dagger$School of Physics \& CTP }
\centerline{\it Seoul National University \& CTP} \centerline{\it
Seoul 151-742, Korea }
\bigskip\bigskip

\abstract{In this paper we construct the 2+1 effective theory of
the light states in D2/D4-brane system in the context of
noncommutative Yang-Mills theory. This effective theory is
noncommutative and tachyonic, however, it is not taking the form
of an Abelian Higgs model as naively expected. We solve the
classical solutions of the effective theory which are nicely
corresponding to different states during the tachyon condensation
process of the dissolution of D2-brane into D4-brane. We also find
that if the expected stable self-dual D0/D4 configuration as the
unit-winding vortex exists, it would be highly calibrated in the
effective theory and be out of the reach of the analytic
solutions.}


\end{titlepage}

\setcounter{footnote}{0}

\section{Introduction}

    Noncommutative Yang-Mills Theory is extensively studied in a classic paper
\cite{SW} by Seiberg and Witten in the context of string theory as the
effective theory of D-brane world volume with the large B-field on it. In that
paper, the D0-brane as a noncommutative instanton on the D4-brane world volume
is analyzed, the perturbative string spectrum shows the existence of the towers
of light states for large noncommutativity. It has been argued that these large
number of light states could be interpretated as the fluctuation modes around
the noncommutative instanton. However, there seems no good formalism to
describe the dynamics of these light states including tachyon except in the
context of noncommutative solitons\cite{GMS1,GMS2,AGMS,NekGro,Poly} as
described below. The existence of towers of light states has been shown in
other Dp/Dq systems with appropriate large B fields in \cite{CIMM1,w0p}, see
also \cite{park}.

Recently the unstable solitons(so called fluxons) in $2+1$ noncommutative
Yang-Mills theory(NCYM) have been constructed in \cite{AGMS}, see also
\cite{NekGro,Kraus,park,Terashima,Hashimoto}. Surprisingly the fluctuations
around the soliton have the same towers of light states spectrum as the ones
obtained from open string excitations of D0/D2 in the large noncommutative
limit. In a recent paper \cite{Rang}, the light states spectrum of D0/D4 open
string with appropriate large B field has been shown to coincide with the
fluctuation spectrum of soliton solution in $4+1$ NCYM. In fact, from the
discussion in \cite{AGMS,Rang}, it is easy to generalize this coincidence to
D0/D6 and D0/D8 cases (see also \cite{Fujii} for more detailed discussions).
The other approach to understand the dynamics of the light states and tachyon
is from on-shell open string amplitude (see \cite{Narain, CIMM2}). It has been
shown the low energy effective action of tachyon is a noncommutative Yang-Mills
with tachyon covariant coupling. The kinetic term of tachyon is found to be
independent of the transverse directions. And there is a damping factor which
shows the noncommutative soliton configuration.  All of these discussions imply
a possible way to understand the large number of light states and the tachyon
condensation in the context of NCYM.

In this paper, we discuss the D2/D4-brane system with B field in
the framework of NCYM. The effective theory of the Dp/D(p+2)-brane
system with B-field, especially the quartic tachyon potential has
been discussed in the commutative description \cite{Narain}.
However, by taking the advantage of easy construction of the
noncommutative solitons which are the co-dimensional D-branes in
Sen's tachyon condensation program
\cite{Sen,HKL0,Seiberg,GMS2,AGMS}, we re-consider the D2/D4 system
in NCYM context and construct the possible solitonic states of the
effective theory dictated by the tachyon potential. The reason for
the effective field theory to be a reliable description of the
tachyon condensation is that the mass of the tachyon is very small
and the stringy effect can be suppressed in the large B-field
limit.

We will start from the D4-brane worldvolume theory and  construct
unstable D2-brane as a fluxon from the point of view of the
transverse directions. Similar to the D0/D2 fluxon, there exist
light fluctuating modes including tachyon which are governed by a
2+1 noncommutative effective theory. One could expect that such an
effective action should be reduced to the noncommutative Abelian
Higgs model with tachyon as a bi-fundamental Higgs field,
disregarding the towers of scalars. The tachyon condensation then
could result in a noncommutative vortex of nontrivial winding
which can be identified as the stable D0-brane on D4-brane world
volume as suggested in \cite{AGMS}. Unfortunately this vortex
configuration turns out to be highly calibrated and is out of the
reach of the analytic solutions in our effective theory
description. The main reason for this complication is that the
tachyon does not have the canonical bi-fundamental scalar
couplings so that the usual partial isometry techniques are
inapplicable here.

The paper is organized as follow: in section 2, the results in
D0/D2 case are briefly reviewed; in section 3, the effective
theory of the light states and the tachyon is given in the case of
D2/D4; in section 4, the classical solutions of the effective
theory are worked out and the possible existence of the
unit-winding vortex solution is discussed; some remarks and
conclusion are in section5. For completeness we give the tachyon
spectrum analysis of D0/D4-brane system in the Appendix.

\section{Light States Spectrum of D0/D2-Brane System}

In this section, we give a brief review of the light states spectrum of
D0/D2-brane system to set up the notations for our discussion on D2/D4-branes.
The detailed discussions can be found in \cite{AGMS}.

\subsection{Soliton solution and effective theory for light states}
The action of the two-dimensional noncommutative Yang-Mills theory can be
recast into a form of Matrix Model(see also \cite{Seiberg}.):
\bea
S&=&-\frac{1}{4g_{YM}^2}\int d^3x\; F_{\mu\nu}F^{\mu\nu} \nonumber \\
 &=&\frac{2\pi\theta}{g_{YM}^2}\int dt \Tr(-\p_t\overline{C}\p_t C
 +\frac{1}{2}([C, \overline{C}]+\frac{1}{\theta})^2)\;.
\eea

Here, following the notation in \cite{AGMS}
\bea
C&=&a^\dagger-iA_z \nonumber \\
\overline{C}& =&a+iA_{\zb}
\eea
with $[a, a^\dagger]=\frac{1}{\theta}$.

The equation of motion is
\be
\p^2_tC=[C, [C, \overline{C}]]\;,
\ee
and the Gauss law constraint for $A_0=0$ is
\be
\label{gauss}
[\Cb,\p_tC]+[C,\p_t\Cb]=0\;.
\ee

   The ground state solution is $C=a^\dagger$ corresponding to a D2-brane.
Moreover, there are static soliton solutions corresponding to m D0-branes
on the D2-brane worldvolume:
\be
C=(S^\dagger)^m a^\dagger S^m+\sum^{m-1}_{i=0}c^i|i><i|
\ee
where $c^i$'s are the constant moduli parameterizing the locations of the
D0-branes, and the "shift" operator $S$ is defined as
\be
\label{shift}
SS^\dagger=1, \hspace{5ex} S^\dagger S=1-P_{m-1}\;,
\ee
with $P_{m-1}\equiv\sum^{m-1}_{i=0}|i><i|$.

In this paper, we focus on the $m=1$ case. Written in the form of matrix,
$$
\label{ss}
C=\left(\begin{array}{cc}
c^0&0 \\
0&a^\dagger
\end{array}
\right)\;.
$$
For simplicity, we take $c^0=0$.

Given the solution for the soliton configuration of (\ref{ss}), we
can investigate the fluctuations around the solution. Out of the
surprise, the spectrum of the fluctuations agree exactly with the
spectrum of light states existing in the D0/D2 system with
appropriate large NS B-field\footnote{In fact, there is a minor
mismatch: $m^2=1/\t$ mode is absent in the fluctuation spectrum
but this mode exists in the perturbative string spectrum. There is
also a mysterious open string scale factor discrepancy between
these two spectra \cite{AGMS}.} \cite{SW,CIMM1,AGMS}. These light
states are the collective excitations around the D0-brane soliton,
taking the form of
$$
\delta C=\left(\begin{array}{cc}
A&W\\
\bar{T}&D
\end{array}
\right)\;,
$$
where the field $A$ is the fluctuations of D0-D0 open strings, while the field
T and W are D0-D2 excitations, and the field D is D2-D2 excitations which
should be integrated out to get the effective potential for the tachyon
describing the dissolution of D0-brane. In components,
\be
\label{comp1}
T=\sum_{k=0} T_k|0><k+1|\;, \qquad W=\sum_{k=0} W_k |0><k+1|\;, \qquad
A=A_0|0><0|\;.
\ee
Note that $T_0$ mode is tachyonic.

  In evaluating the effective potential
\be
V={1\over 2}([C,\Cb]+{1\over \t})^2
\ee
we will restricted only to the sector
\be
\label{comp2}
D=\sum_{k=0} D_{k+1,k}|k+1><k|\;,
\ee
for the reasons as pointed out in \cite{AGMS}. After some lengthy calculation,
the effective potential takes the form of the sum of complete squares
\bea
\label{V}
2V &=& M^2_{-1}+M^2_0+(N_0 \Nb_0+c.c.)
\nonumber
\\
&+& \sum_{k=1}M^2_k + \sum_{k=1}(N_k \Nb_k+c.c.)
\nonumber
\\
&+& \sum_{k=1}(Q_k \Qb_k +c.c.)
\nonumber
\\
&+& \sum_{i=1}\sum_{k=1}(R_{i,k}\Rb_{i,k}+c.c.)\;.
\eea

where
\bea
\label{first}
M_k&=&E_k \Eb_k - \Eb_{k+1}E_{k+1}+{1\over \t}
-{k+2 \over 2k+3} \Jb_{k+1} J_{k+1} + {k-1 \over 2k-1}
\Jb_{k-1} J_{k-1}\;,
\\
N_k&=& \sqrt{1 \over 2k+1}J_k(\sqrt{k+1}\Eb_k+\sqrt{k}\Eb_{k+1})
\nonumber
\\
&-&(\sqrt{k+2\over 2k+3}\Ab J_{k+1}+\sqrt{k-1\over 2k-1}A J_{k-1})\;,
\\
M_{-1}&=&-T_0 \Tb_0 +{1\over \t} +[A_0,\Ab_0]
+\sum_{k=1}{1\over 2k+1} J_k \Jb_k\;,
\\
M_0&=&\Tb_0 T_0 - E_1 \Eb_1 +{1\over \t}-{2\over 3}\Jb_1 J_1\;,
\\
N_0&=&A_0T_0-\sqrt{2\over 3}\Ab_0J_1\;,
\\
Q_k&=&-(\sqrt{{2(k+2)\over 3(2k+3)}} \Jb_1 J_{k+1}
+\sqrt{{k-1 \over 2k-1}} \Tb_0 J_{k-1})\;,
\\
\label{final}
R_{i,k}&=&-\sqrt{{(i+2)(i+k+2)\over (2i+3)(2i+2k+3)}} \Jb_{i+1} J_{i+k+1}
\nonumber
\\
&-&\sqrt{{(i-1)(i+k-1)\over (2i-1)(2i+2k-1)}}\Jb_{i-1} J_{i+k-1}\;,
\eea
and $E_k$'s are defined as
\be
E_k \equiv D_{k,k-1}+\sqrt{k\over \t}\;.
\ee
In the above we have used the Gauss Law constraint
\be
0={1\over \sqrt{2k+1}}(\sqrt{k}W_k+\sqrt{k+1}T_{k+1})\;,
\ee
and define a new scalar
\be
J_k={1\over \sqrt{2k+1}}(\sqrt{k+1}W_{k-1}-\sqrt{k}T_{k+1})\;,
\ee
so that
\be
W_{k-1}=\sqrt{{k+1 \over 2k+1}} J_k\;, \qquad T_{k+1}=-\sqrt
{{k\over 2k+1}}J_k
\ee
with $T_1=J_0=0$ and $k=1,2,3,\cdots$.

\subsection{The spectrum and tachyon condensation}
Some remarks are in order about the above results. Firstly, In the case of
D0/D2, the fluctuations around unstable soliton configuration are just complex
excitations, and they can be described by quantum mechanics. We will show in
the following section in the case of D2/D4, the fluctuations have to be
governed by 2+1 dimensional noncommutative field theory. We have retained the
order of the fields in (\ref{V}) so that the result can be directly carried to
its noncommutative version for the D2/D4-brane system.

Another interesting point is that what is the fate of the fluctuations after
tachyon condensation. As discussed in \cite{AGMS}, there exist two vacua: one
is unstable vacuum for the undissolved D0/D2, and the other is the stable
vacuum for the dissolved D0 in D2. Thanks to the sum of complete squares of
(\ref{V}), the stable vacuum is at the minimum $V=0$ with
\be
\label{stablev}
M_{-1}=M_0=M_k=N_0=N_k=Q_k=R_{i,k}=0\;,
\ee
which can be solved by
\be
T_0=\sqrt{{1\over \t}}\;, \qquad E_k=\sqrt{{k+1\over\t}}\;,
\ee
and the other fields are set to zero. On the other hand, the unstable vacuum of
$D0/D2$-brane is at
\be
T=W=D=A=0\;,
\ee
with $2V={1\over \t^2}$.

It's easy to read out the mass spectrum of excitations from the quadratic terms
in (\ref{V}): Around the unstable vacuum, the worldvolume gauge field $A_0$ of
D0-brane is massless, while the tachyon $T_0$ has the mass-square
$-\frac{1}{\theta}$, and the complex massive scalar $J_k$'s has the mass-square
$\frac{2k+1}{\theta}$, which are light for large $\t$ and are exactly the
masses of light states of D0-D2 open string in the zero slope
limit\cite{CIMM1,SW}. Around the stable vacuum, the excitations $A, J_k$ get
extra masses from the tachyon condensation. In fact, the relevant quadratic
terms of the action turns out to be\footnote{We neglect the mass terms for
$D_{k,k-1}$ scalars since they are the D2-D2 open string fluctuations which are
not localized around the D0-brane, and should be integrated out for the
effective theory.}
\be
\frac{1}{\t} A_0 \Ab_0 + \frac{1}{\t} \sum_{k=1}
\frac{(k+1+\sqrt{k(k+2)})^2}{2k+1} J_k \Jb_k \label{masscon}
\ee
Obviously the excitation $A_0$ get mass implying the dissolution of the
D0-brane is described by Higgs mechanism, and $J_k$ get a little bigger mass
damping the fluctuations after tachyon condensation. This tells that the
collective fluctuations do not decouple but are damped after the tachyon
condensation.

\section{Effective Theory of Light States of D2/D4-Brane System}

In this section, we would like to discuss  the effective noncommutative field
theory of light states in D2/D4-brane system. We will start from the D4-brane
world volume action and construct the D2-branes as the unstable soliton. This
is quite similar to the construction of D0/D2. The key difference is that we
will obtain an effective noncommutative field theory which governs the dynamics
of the fluctuations.

\subsection{The world volume theory of a D4-brane}
The world volume action of a D4-brane with a B field background on it is a
$(4+1)$-dimensional noncommutative Yang-Mills theory, for simplicity B takes a
canonical form, i.e.
\be
\label{canon}
B_{\mu\nu}=\left(\ba{cccc}
0&B_{12}& & \\
-B_{12}&0& &\\
 & &0& B_{34}\\
 & &-B_{34}&0
\ea\right).
\ee
As usual, $B_{12}$($B_{34}$) induces noncommutativity in
$1,2$($3,4$)-directions with parameter $\t_1$($\t_2$).

Then the action can be written in the form
\bea
\label{act}
S&=&{-1\over 4 g_{YM}^2}\int d^5x\; F_{\mu \nu}F^{\mu \nu}
\nonumber
\\
\label{24action}
 &=& -\frac{4\pi^2\theta_1\theta_2}{g^2_{YM}}\int dt
\Tr(-\p_t\overline{Y}\p_tY
-\p_t\overline{Z}\p_tZ \nonumber \\
& & +\12 ([Y, \overline{Y}]+\frac{1}{\theta_1})^2 +\12 ([Z, \overline{Z}]+\frac{1}{\theta_2})^2 \nonumber \\
& & +[\overline{Y}, \overline{Z}][Z, Y]+[Y, \overline{Z}][Z, \overline{Y}])
\eea
where
\be
Y=a_y^\dagger-iA_y(y,\overline{y},z,\overline{z}), \hspace{5ex}
Z=a_z^\dagger-iA_z(y,\overline{y},z,\overline{z})
\ee
with $[a_{y,z}, a^\dagger_{y,z}]=\frac{1}{\theta_{1,2}}$. As be will be shown
that the cross terms between $Y$ and $Z$ will give the correct kinetic terms of
the collective excitations in the effective theory of D2/D4 system.

We are only interested in the classical static solution, so the equations of motion
are
\be
[Y, [Y, \overline{Y}]]+[\overline{Z}, [Y, \overline{Z}]]+[Z, [Y,
\overline{Z}]]=0\;,
\ee
\be
[Z, [Z, \overline{Z}]]+[[Y, Z],\overline{Y}]]+[[\overline{Y},
Z],\overline{Y}]=0\;,
\ee
and the Gauss Law constraints are just the same as (\ref{gauss}).

The ground state(zero energy) solution corresponding a D4-brane is
\be
Y=a^\dagger_y\otimes I_z\;, \hspace{3ex} Z=I_y\otimes a^\dagger_z\;,
\ee
where $I_{y(z)}$ is the identity operator in the subspace of $1,2(3,4)$
noncommutative plane, which indicates the infinite extension along the
corresponding directions.

\subsection{Effective noncommutative field theory }
The soliton solution of (\ref{act}) corresponding to a D2/D4-brane
configuration is
\be
\label{24}
Y=C_y\otimes P_{0z}\;,  \hspace{3ex} Z=I_y\otimes S_z^\dagger a^\dagger_z
S_z\;.
\ee
Here $C_y$ can be thought as the background field coupled to the fluctuating
modes. Note that the solution corresponding to the D2/D4 with zero background
$C_y=0$ has the zero-point energy
\be
\label{0point}
E_0 = {2\pi^2 \over g^2_{YM}}({\t_1\over \t_2}+{\t_2\over \t_1})
\Tr_y{I_y}=E_{D0} \Tr_y I_y.
\ee
the trace $\Tr_y$ is taken over $1,2$-noncommutative plane, and $E_{D0}$ is the
energy of a D0-brane on D4-brane as given in the Appendix. The form of the
zero-point energy indicates the underlying D2-brane which can be viewed as the
bound state of infinite D0-branes on the $1,2$-plane.

    From the D4-brane point of view, the solution (\ref{24}) gives an unstable
soliton which is just a D2-brane extending along the $1,2$-directions but
localizing in the $3,4$-directions. Therefore, we can obtain a noncommutative
effective field theory for the collective excitations living on the 1,2-plane
by treating the solution (\ref{24}) as a D0/D2-brane system in the
$3,4$-directions.  The collective excitations as the fluctuations around the
configuration (\ref{24}) can be parameterized as
\be
\delta Z=\left(\ba{cc}
A(y,\overline{y})& W(y, \overline{y})\\
\overline{T}(y, \overline{y})& D(y, \overline{y}) \ea\right).
\ee
Note that all fluctuations now are the fields depending on the $1,2$
coordinates $y, \overline{y}$ and their products should be understood as
star-products. Similar to the D0/D2 system, $A$ corresponds to the 2-2 string excitations,
$T$ and $W$ corresponds to the 2-4 string excitations and $D$ to the 4-4 string excitations,
which
should be integrated out in the effective theory. Again all the modes becomes
light states of the string theory in the large noncommutativity
limit\cite{SW,CIMM1}.

   To obtain the effective theory, we need to expand $A,W,T,D$ into components
as done in (\ref{comp1}) and (\ref{comp2}) but now each component field is
function of $y$ and $\overline{y}$. Again we keep on the branch of $D$ as
given by (\ref{comp2}).

First the term ${1\over 2}\Tr([Z, \overline{Z}]+1/\theta_2)^2$ in
(\ref{24action}) gives part of the effective action $S_1$ for the fluctuation
fields as following
\bea
\label{U}
2S_1&=&\Tr_y ({1\over \t_2^2}+\; M^2_{-1}+M^2_0+(N_0 \Nb_0+c.c.)
\nonumber
\\
&+& \sum_{k=1}M^2_k + \sum_{k=1}(N_k \Nb_k+c.c.)
\nonumber
\\
&+& \sum_{k=1}(Q_k \Qb_k +c.c.)
\nonumber
\\
&+& \sum_{i=1}\sum_{k=1}(R_{i,k}\Rb_{i,k}+c.c.) \;)\;.
\eea
where the fields $M,N,Q,R$ are the same as the ones defined in (\ref{first}) to
(\ref{final}) with $\t$ replaced by $\t_2$. As remarked before, although these
fields are noncommuting, we just directly carry (\ref{V}) of D0/D2 system to
(\ref{U}) because we have retained the order of the fields in the expression of
(\ref{V}).

   On the other hand, the rest of the action (\ref{24action}) will give another
part of the effective action with background field couplings
\bea
S_2 &=&\Tr_y(\; \12 ([C_y, \overline{C}_y]+\frac{1}{\theta_1})^2)+\Tr_y
\Tr_z(\; [\overline{C}_y\otimes P_{0z}, \delta\overline{Z}][\delta Z,C_y\otimes
P_{0z} ]
\nonumber\\
&+&[C_y\otimes P_{0z},\delta\overline{Z}][\delta Z, \overline{C}_y \otimes
P_{0z} ]\;)\;.
\eea
Here we have suppressed the fluctuation $\delta Y$ which may be relevant for
the complete light states spectrum as discussed in the Appendix for the
D0/D4-brane system.

The total energy for the effective theory of the light states is given by
\be
\label{energy}
E={4\pi^2\t_1\t_2 \over g^2_{YM}}(S_1+S_2)\;.
\ee

\section{Tachyon Condensation of D2/D4-Brane System}
In this section we will discuss the tachyon condensation of the unstable
D2-brane from the effective field theory we obtained in the last section.

\subsection{Tachyon condensation in an Abelian Higgs model?}

From the 2+1 effective field theory action $S_1+S_2$ the equations of motion can
be obtained as usual. The tachyon field and the scalar fields coming from the
2-4 string are bi-fundamental complex scalars, one would expect that the
effective action is an Abelian Higgs model which has the noncommutative vortex
solution of nontrivial winding \cite{Wadia,Bak,Kraus} which could be identified
as a D0-brane. It turns out this is not the case since their explicit forms of
couplings with gauge fields are not of the form of bi-fundamental or
fundamental scalar coupling if we keep the potential in the Higgs form.

To be more explicit, we have two gauge fields: one is the gauge field $A_0$
from 2-2 string which reflects the dynamics of the underlying D2-brane, and the
other is the background gauge field $C_y$ which represents the more general
D2/D4-brane configurations and is turned off in the following discussions. For
comparison, let us recall the form of the covariant coupling of a
bi-fundamental scalar $\phi$
\be
\label{bifun}
\Tr(\;\overline{C}_1\phi\overline{\phi}C_1+C_1\phi\overline{\phi}\overline{C}_1
+\overline{C}_2\phi\overline{\phi}C_2+C_2\phi\overline{\phi}\overline{C}_2-2C_1\phi
\overline{C}_2\overline{\phi}-2C_2\overline{\phi}\overline{C}_1\phi\;).
\ee
with $C_1=a^\dagger-iA_1$ and $C_2=a^\dagger-iA_2$.

On the other hand, the relevant action for tachyon condensation from $S_1+S_2$
by setting $J_k=0$ is
\bea
\label{Seff}
S_{eff}& &=\Tr_y (\;\12 ([C_y,
\overline{C}_y]+\frac{1}{\theta_1})^2+{1\over2}(-T_0\overline{T}_0+{1\over
\t_2}+[A_0, \overline{A}_0])^2+A_0T_0\overline{T}_0\overline{A}_0
\nonumber \\
& &+C_yT_0\overline{T}_0\overline{C}_y+\overline{T}_0C_y\overline{C}_yT_0 +
+[\overline{C}_y, \overline{A_0}][A_0, C_y]+ [C_y, \overline{A_0}][A_0,
\overline{C}_y]\;)\;.
\eea
Obviously the gauge fields $C_y$ and $A_0$ have the correct kinetic terms; the
tachyon $T$ has the usual Higgs potential but its kinetic term is not in the
canonical form of (\ref{bifun}). Therefore the effective theory (\ref{Seff}) is
not an Abelian Higgs model as naively expected.

\subsection{The classical solutions and tachyon condensation}
In the following we will solve the classical solutions of the effective theory
which are relevant to the tachyon condensation of the D2/D4-brane system.

From the effective action (\ref{Seff}) we derive the equations of motion for
$T_0$, $A_0$ and $C_y$ respectively:
\bea
\label{eqm1}
0&=&(\;C_y\overline{C}_y+\overline{C}_yC_y-T_0\overline{T}_0+{1\over\t_2}-[A_0,\overline{A}_0]
+\overline{A}_0A_0\;)T_0\;,
\\
0&=&[[A_0,C_y],\overline{C}_y]+[[A_0,\overline{C}_y],C_y]-
[-T_0\overline{T}_0+[A_0,\overline{A}_0],A_0]-A_0T_0\overline{T}_0\;,
\\
\label{eqm3}
0&=&[[C_y,\overline{C}_y],C_y]+[\overline{A}_0,[A_0,C_y]]+
[[C_y,\overline{A}_0], A_0]+C_yT_0\overline{T}_0+T_0\overline{T}_0C_y\;.
\eea

  There are many solutions of the equations of motion by simple inspection and by
using the solution generating method based on partial isometry
\cite{Kraus}. We will discuss the relevant ones which can be
interpreted as the initial, intermediate and final states of the
tachyon condensation. We also discuss the solutions as the
possible candidates for D0/D4, however, it turns out that none of
them matches the energy spectrum of D0/D4 given in the Appendix.

   The first interesting configuration is the stable vacuum known as the
"nothing state" representing the complete dissolution of the D2-brane fluxon
into the D4-brane. By definition all the gauge fields should be set to zero and
tachyon vev reach the bottom of its potential, thus
\be
\label{nothing}
C_y=0\;, \qquad T_0={\frac{1}{\sqrt{\t_2}}}\;,\qquad A_0=0\;,
\ee
which indeed solves the equations of motion. Unlike the usual stable ground
state of tachyon condensation, this configuration has the energy
$E_{nothing}={2\pi^2 \theta_2 \over g^2_{YM} \theta_1}\Tr_y I_y$ which should
be subtracted out from the energy of the other configurations. Note that
$E_{nothing}<E_0$ of (\ref{0point}) as a consistency check, also that
$E_{nothing}$ is not symmetric with respect to the exchange of $\theta_{1,2}$
which can be understood as the tachyon condensation relic of the localized
D2-brane breaking the rotational symmetry.

    By using the solution generating method of the partial isometry, one can
generate new solutions with respect to the above solutions. For example,
from the nothing state, we will have the new solution
\be
\label{nt}
C_y=0\;,\qquad T_0={1\over \sqrt{\t_2}}(1-P_{0y})\;,\qquad  A_0=\alpha_0 P_{0y}\;
\ee
where $\alpha_0$ is an arbitrary constant. This solution has the
$\alpha_0$-independent finite net energy $\Delta E={2\pi^2 \t_1\over
g^2_{YM}\t_2}$ with respect to the $E_{nothing}$ of the nothing state. This
configuration represents the dissolution of the D2-brane into a Gaussian bump
carrying no magnetic flux because $[A_0,\overline{A}_0]=0$. Moreover, it has a
tachyonic mode with mass-squared ${-1\over\t_2}$ indicating the tendency of the
complete dissolution of the D2-brane. This is in contrast to the D0/D4-brane
configuration as analyzed in the Appendix, which carries flux or instanton
charge and is stable if $\t_1=\t_2$.

In principle we would think the corresponding D0/D4 configuration
as given in the Appendix will appear near the tachyon vacuum such
that the tachyon vev will take the form as in  (\ref{nothing}) or
(\ref{nt}). One may then wonder if there exist a modified solution
of (\ref{nothing}) or (\ref{nt}) with nontrivial $A_0$
configuration such that one can identify it as the D0/D4
configuration with the correct net energy $E_{D0}$?

\subsubsection{Vortex solution of unit winding?}
   One of the possible candidates for the $D0/D4$ as suggested in \cite{AGMS} and
\cite{Bak}, also in the Appendix is the vortex solution of
unit-winding which takes the following ansatz form
\be
\label{ansatz1}
T_0=S_y \equiv \sum_{k=0} |k><k+1|\;\; or \;\; S^{\dagger}_y \;
\qquad C_y=\beta S_y^{\dagger} a_y^{\dagger} S_y; \qquad
A_0=\alpha S^{\dagger}_y a^{\dagger}_y S_y\;,
\ee
where $|\alpha|^2+|\beta|^2=1$ from the requirement of the unit
winding. If our tachyon is a bi-fundamental scalar in the
Abelian-Higgs model, then (\ref{ansatz1}) will solve the equations
of motion, and the vortex solution can be interpreted as the
D0/D4.  However, this is not the solution since our effective
theory is not a bi-fundamental Abelian-Higgs model.

  We should then try more general ansatz for the tachyon as
\be
T_0=\sum_{k=0} t_k |k><k+1|\;\; or \;\; \sum_{k=0} w_k |k+1><k|\;.
\ee
Note that if $T_0$ contain both the above sectors, then there is
no consistent truncation to these subspace while solving the
equations of motion.  For the moment we still require the simple
form of the gauge fields as in (\ref{ansatz1}) such that all the
double commutators in (\ref{eqm1})-(\ref{eqm3}) involving only the
gauge fields will vanish.

   Using the ansatz, the equation (\ref{eqm1}) gives
\be
|t_{k-1}|^2={1\over \theta_2}+{1\over
\theta_1}(2k-1+|\alpha|^2(2-k))\;,\;\; k=1,2,3,...
\ee
Similarly for the case of $w_k$.  However, the equation
(\ref{eqm3}) gives
\be
|t_k|^2+|t_{k-1}|^2=0\;, \;\; k=1,2,3,...
\ee
It is clear that the two conditions contradict so that there is no
solution for this ansatz.

    The most generic ansatz is to also assume the general form for
the gauge fields as
\be
\label{ansatz3}
C_y={1\over \sqrt{\theta_1}}\sum_{k=0} c_k |k+1><k|\;, \qquad
A_0={1\over \sqrt{\theta_1}}\sum_{k=0} a_k |k+1><k|\;,
\ee
and the unit winding requirement
\be
\label{unit3} \{([A_0,\bar{A}_0]+{1\over
\theta_1})+([C_y,\bar{C}_y]+{1\over \theta_1})\}={1\over
\theta_1}|0><0|
\ee
will constrain $a_k$s and $c_k$s by
\bea
|a_0|^2&+&|c_0|^2=1\;,
\\
|a_k|^2&+&|c_k|^2=|a_{k-1}|^2+|c_{k-1}|^2+2\;, \;\; k=1,2,3,...
\eea
In (\ref{unit3}) we have assumed the unit flux is completely
localized in the $|0><0|$ sector, more general situation with the
unit flux scattered in different sectors will lead to more
complicated constraints on $a_k$ and $c_k$.

The ansatz (\ref{ansatz3}) makes the double commutators in the
equations of motion non-vanishing and leads to the very
complicated coupled {\it cubic} recursive equations for $a_k$,
$c_k$ and $c_k$'s so that one cannot obtain the analytic
solutions. However, this does not rule out the possible existence
of a highly calibrated unit-winding solution which is just out of
the reach of the analytic solutions.

\subsubsection{The other non-D0/D4 solutions}

  It is also interesting to study the configurations near the "top" of the tachyon
potential. The configuration right on the top of the potential is
\be
\label{top}
C_y=0,\qquad T_0=0, \qquad A_0=a^\dagger_y
\ee
which also solves the equations of motion. Note that the tachyon
vev is zero and the gauge field $A_0$ representing the D2-brane is
turned on\footnote{ For more general consideration, one can also
turns on the background gauge field $C_y$ in (\ref{top}) such that
$C_y= a^{\dagger}$. This kind of configurations have the energy
$E={2\pi^2\over g^2_{YM}}({2\t_2\over \t_1}+(\sqrt{{\t_2 \over
\t_1}}-\sqrt{{\t_1\over \t_2}})^2) \Tr_y I_y$.}. This
configuration has the energy $E=E_{nothing}+{2\pi^2 \over
g^2_{YM}}(\sqrt{\frac{\t_1}{\t_2}}- \sqrt{\frac{\t_2}{\t_1}})^2\;
\Tr_y I_y$ which is always larger than the energy $E_{nothing}$ as
expected unless $\t_1=\t_2$.  Note that the energy bears similar
form of the one for D0/D4-brane system \cite{AGMS} except that
there is an $\Tr_y$ factor indicating the infinite number of the
constituent D0-branes.

To check the stability of (\ref{top}) we need to look into its fluctuation
spectrum which takes the following form
\be
({1\over \t_2}-{1\over\t_1})\Tr_y (-\delta T_0 \delta
\overline{T}_0+[a^+_y,\delta A_0][\delta \overline{A}_0,a_y])+a^+_y \delta T_0
\delta \overline{T}_0 a_y \;.
\ee
The lowest mode of $\delta T_0$ has the mass-square
${2\over\t_1}-{1\over\t_2}$ and the modes of $\delta A_0$ has the mass-square
proportional to ${1\over\t_2}-{1\over\t_1}$ so that there is a window $2\t_2
\ge \t_1 \ge \t_2$ for the absence of the tachyon. There is no clear physical
reason to explain the  existence of such a window which implies the possibility of stable
configurations. Outside this window, the configuration is unstable and will
condense toward the nothing state.

  Finally there is a one-flux solution near the top of the potential which
however can not be generated from (\ref{top}) by the partial isometry, taking
the form
\be
C_y=0\;, \qquad T_0={1\over \sqrt{\t_2}}P_{0y}\;,
\qquad A_0=S^{\dagger}_ya^{\dagger}_yS_y
\ee
with the negative net energy $\delta E=-{2\pi^2 \over
g^2_{YM}}(\sqrt{\frac{\t_1}{\t_2}}- \sqrt{\frac{\t_2}{\t_1}})^2$
with respect to the energy of the configuration (\ref{top}) indicating that a
Gaussian flux bump is created on D2/D4 due to the tachyon condensation. This
flux bump is nothing but a hole left by an evacuated D0-brane out of D2-brane
since it has $-1$ flux with respect to the flux number of the configuration
(\ref{top}). Moreover, this configuration is unstable as one can see from its
fluctuation spectrum so that the system will continue to condense and lower its
energy until it reaches the nothing state.

 The other solutions with the similar kind of properties of the above flux
bump are listed below:
\begin{equation}
\ba{llll} 1)&
C_y=a^{\dagger}_y\;,&\qquad T_0=0\;,& \qquad A_0=0\;,\\
2)&
C_y=a^{\dagger}_y\;,&\qquad T_0=0\;, &\qquad A_0=\kappa a^{\dagger}_y\;,\\
 & \mbox{where $\kappa$ is a constant.}& & \\
3)&
C_y=S^{\dagger}_ya^{\dagger}_yS_y\;, & \qquad T_0=0\;, &\qquad A_0=0\;,\\
4) & C_y=S^{\dagger}_ya^{\dagger}_yS_y\;, &\qquad T_0={1\over
\sqrt{\t_2}}P_{0y}\;,&
\qquad A_0=0\;, \\
5) & C_y=S^{\dagger}_ya^{\dagger}_yS_y\;, &\qquad T_0=0\;,&
 \qquad A_0=S^{\dagger}_ya^{\dagger}_yS_y\;, \\
6)& C_y=S^{\dagger}_ya^{\dagger}_yS_y\;,& \qquad T_0={1\over
\sqrt{\t_2}}P_{0y}\;,& \qquad A_0=S^{\dagger}_ya^{\dagger}_yS_y \;.\ea
\end{equation}

\section{Conclusion}

   Starting from the 4+1 NCYM we have constructed the noncommutative effective
theory for the light states of D2/D4-brane system. Before
examining the details of the effective theory, it is quite
tempting to expect a noncommutative Abelian Higgs model such that
its vortex solution may be interpreted as some stable D0/D4
configurations given in the Appendix. Unfortunately, this is not
the story we find above. In fact, the effective action is not an
Abelian Higgs model so that if the unit-winding D0/D4 solution
exists it would be out of the reach of the analytic solutions.
Despite that, some of the solutions of the effective theory have
nice interpretations in the process of tachyon condensation
describing the complete dissolution of D2 into D4.

   The similar situation happens in the Dp-anti-Dp-brane system where it is
difficult to construct an effective Abelian Higgs theory and also a tachyonic
funnel solution as the noncommutative vortex in the context of NCYM at the
large B-field limit \cite{ddbar}. This may be seen as the limitations of the
NCYM as the effective theory in describing some configurations made of highly
calibrated collective modes in the tachyon condensation. This also indicates
the need of carefully choosing the variables of the effective theory to
describe some nontrivial configurations; a famous example is how to describe
closed string around the tachyon vacuum \cite{closed}. A direct study of the
tachyon condensation of these systems in the off-shell string theory should be
pursued to understand the complications due to the towers of both light and
stringy states which are completely neglected in our current treatment.

 A final side remark is about some novel solutions of the
equations of motion (3.4) and (3.5), which has the following form similar to
D0/D4 configuration given in the Appendix:
\bea
Y&=& S^{\dagger m}_y a^\dagger_y S^m_y \otimes P_{kz} \nonumber \\
Z&=& P_{ly} \otimes S^{\dagger n}_z a^\dagger_z S^n_z
\eea
where $k\leq n$ and $ l\leq  m$. In fact, here the $Y$ and $Z$ decouple
completely. The energy of such solutions is
$$
E=\frac{2\pi^2}{g^2_{YM}}(\frac{\theta_2}{\theta_1}mk+\frac{\theta_1}{\theta_2}nl)
$$
which is finite. The interpretation of these solutions in string theory is
unclear at this moment. Some of them could be the solutions $T^{\dagger
n}a^\dagger_m T^n$ as in \cite{AGMS} and be thought as the intersecting
localized D2-branes as considered in \cite{tseng}. It deserves more study of
these configurations.

\vskip 2cm

{\bf Acknowledgements}

FLL is grateful to the hospitality of ICTP at Trieste and KIAS at Seoul where
part of the work was carried out, and he would like to thank Kimyeong Lee,
Jeong-Hyuck Park and Piljin Yi for helpful discussions. FLL is supported by
BK-21 Initiative in Physics (SNU-Project 2).

\section{Appendix: The light states spectrum of D0/D4-brane}
The soliton solution of
(\ref{act}) corresponding to a D0/D4-brane is given by
\be
Y=S^{\dagger}_y a^{\dagger}_y S_y \otimes P_{0z}\;, \qquad Z=P_{0y}\otimes
S^{\dagger}_z a^{\dagger}_z S_z\;.
\ee
where $S_{y(z)}$ is the shift operators in the $1,2(3,4)$-directions as defined
in (\ref{shift}), and $P_{0y(z)}$ is the corresponding zero-sector projection
operator. The energy of this configuration is finite and equals to
$E_{D0}\equiv{2\pi^2 \over g^2_{YM}}({\t_1\over\t2}+{\t_2\over\t_1})$. In
\cite{AGMS} it has been conjectured that the fluctuation spectrum contains a
tachyon unless the B field configuration is (anti-)self-dual, i.e. $\t_1=\pm
\t_2$. This fact is also shown in the perturbative string theory calculations
as a BPS condition for preserving the supersymmetry \cite{SW,CIMM1}. In the
following we will demonstrate this explicitly in the context of NCYM.

    The complete fluctuation spectrum is quite complicated, we are only
interested in the $A_0$ and $T_0$ sectors which correspond to the gauge field
and tachyon. Then the fluctuation can be written in the following form
\be
\delta Y=(A_{0y}|0><0|+\Tb_{0y}|1><0|)\otimes P\;, \qquad \delta Z=P \otimes
(A_{0z}|0><0|+\Tb_{0z}|1><0|)\;.
\ee
where $P=(|0><0|+|0><1|+|1><0|)$ is the operator representing the most possible
fluctuation along the transverse directions caused by the $A_0,T_0$ sector due
to the energy conservation. We also assume that we have fluctuations around an
almost self-dual configuration so that we choose the same $P$ to both $\delta
Y$ and $\delta Z$ to maintain the rotational symmetry of the configuration.

   After some calculations by using the above ansatz, the quadratic action for
$A_0$ and $T_0$ is
\be
({1\over \t_2}-{1\over \t_1})|T_{0y}|^2+({1\over \t_1}-{1\over
\t_2})|T_{0z}|^2+ {2\over \t_2} |A_{0y}|^2+{2\over \t_1} |A_{0z}|^2\;,
\ee
where the negative contribution to the tachyon mass comes from the same origin
as in the D0/D2 case, and the positive one from the terms of the commutators
between $Y$ and $Z$.

    It is obvious that there is always a tachyonic mode unless the
configuration is self-dual, i.e. $\t_1=\t_2$. However, the gauge fields become
massive, implying no massless modes in the fluctuation spectrum. In \cite{Rang,
Fujii}, the full spectrums of the scalar fluctuations and the gauge field
fluctuations have been worked out by using the other basis.


\end{document}